# Zero-crossing Shapiro steps in focused-ion-beam-tailored high-$T_c$ superconducting microstructures

Myung-Ho Bae[1], R. C. Dinsmore III[1], M. Sahu[1], Hu-Jong Lee[2], and A. Bezryadin[1]

[1]Department of Physics, University of Illinois at Urbana-Champaign, Urbana, Illinois 61801-3080, USA

[2] Department of Physics, Pohang University of Science and Technology, Pohang 790-784, Republic of Korea

**Abstract:**

Microwave response of S-shaped $Bi_2Sr_2CaCu_2O_{8+x}$ (BI-2212) micron-scale samples, in which the supercurrent was forced to flow perpendicular to the crystal layers, was investigated. A treatment with a focused ion beam allowed us to reduce the plasma frequency down to $f_p$~5 GHz at $T$=0.3 K in naturally stacked Josephson junctions in a crystal. We observed Shapiro steps at frequencies as low as ~5 GHz. Well-developed zero-crossing Shapiro steps were observed at frequencies as low as ~10 GHz. They appeared as constant-voltage plateaus with a non-zero voltage occurring at zero bias current. We confirmed that zero-crossing Shapiro steps in the Bi-2212 stacked junctions can be observed when the irradiated frequency is sufficiently larger than $f_p$. The observed high-order fractional steps in the microwave responses indicate that the interlayer-coupled Bi-2212 Josephson junctions have nonsinusoidal current-phase relation. Based on the temperature dependence of the steps we also showed that the finite slope of the steps is due to the enhancement of the phase diffusion effect.

## I. INTRODUCTION

The weak coupling between the superconducting layers in $Bi_2Sr_2CaCu_2O_{8+x}$ (Bi-2212) and $Tl_2Ba_2Ca_2Cu_3O_{10+x}$ (Tl-2223) results in the formation of the intrinsic Josephson junctions (IJJs) [1] or stacked junctions along the *c* axis of the crystal. For Bi-2212, the superconducting layers in the crystal are separated by 1.2 nm from each other, giving rise to homogeneously and densely stacked junctions. Evidences for the DC Josephson effect in the Bi-2212 IJJs system have been revealed in various observations such as Fraunhofer patterns [2], flux quantization [3], and thermal activation [4]. Furthermore, the recent observations of the phenomenon of macroscopic quantum tunneling [5,6] in the IJJs system [7] may open platform for the quantum engineering based on high-$T_c$ superconductors. Clear Shapiro steps at some specific frequencies satisfying the AC Josephson relation [8,9,10] have recently been observed in the Bi-2212 IJJs system [11,12,13]. The occurrence of Shapiro steps in a Josephson junction [14,15] is explained by a phase locking effect, *i.e.* by a synchronization of the periodic evolution of the phase of the superconducting wave function, caused by the DC bias voltage, and the oscillation of the external electric field associated with the applied microwave (MW) irradiation [14,10]. The steps occur in the current-voltage (*I-V*) curves at discrete voltage values $V_n = nhf/2e$ (*f*; the frequency of the applied microwave signal, *h*; the Planck constant, *e*; the elementary charge, and *n*; the integer index number) [16].

Especially, highly underdamped tunneling in a Josephson junction leads to Shapiro steps that cross the current axis, *i.e.*, zero-crossing Shapiro steps (ZCSS), for an irradiated frequency sufficiently higher than the junction plasma frequency, $f_p$ [17]. The zero-crossing corresponds to a situation when a DC voltage occurs on a layered crystal, irradiated with microwaves, without any DC current bias applied. In Bi-2212 IJJs, with highly underdamped tunneling behavior, ZCSS were observed in the terahertz range [12]. In this case uniform electromagnetic response over the entire stacked junctions was also demonstrated. The observation frequency of ZCSS, however, shifts to the W-band (*i.e.* 75-100 GHz) on the surface junctions, with the interlayer coupling strength weaker than the inner junctions [11]. Stable Shapiro steps including the zero-crossing ones are only observable for frequencies much higher than $f_p$, which is on the order of a few hundred GHz in Bi-2212 IJJs [7,12]. This has made it difficult to study the detailed frequency dependence of the steps in these IJJs.

In this study, we fabricated S-shaped Bi-2212 micron-scale samples by the focused-ion-beam (FIB) milling [18] and studied Shapiro steps in the IJJs imbedded inside the crystal with the MW in the frequency range of 0.5-26 GHz. The plasma frequency of some junctions in our sample was reduced down to as low as ~5 GHz at $T$=0.3 K by weakening of the coupling strength of stacked junctions due to $Ga^+$ irradiation during the FIB process. This fact gave an important advantage: the range of required MW frequencies was strongly reduced. Above $f$~10 GHz, our samples showed clear ZCSS. At $f$=5 GHz, the behavior turned into hysteretic constant-voltage current steps without zero-crossing. Near $f$=1 GHz, Shapiro steps became absent in the hysteretic curves before jumping to the quasiparticle branches in a higher bias range. For $f$=18 GHz, high-order fractional steps were observed, which may indicate that the interlayer-coupled Bi-2212 IJJs is governed by a nonsinusoidal current-phase relation. For $f$=23 GHz, based on the measured temperature dependence of the steps at a fixed power, we also showed that the finite-slope Shapiro steps are due to the enhancement of the phase-diffusion effect.

## II. EXPERIMENT

Single crystals Bi-2212 was prepared in the conventional solid-state-reaction method. A sample with a lateral size of 4.3×5 $\mu$m$^2$ tailored in an S-shaped geometry, indicated by the gray region in the schematics of the lower inset of Fig. 1, was fabricated by the FIB process. We used a relatively high ion-beam current of 3 nA, which corresponded to ~200 pA/$\mu$m$^2$, in the milling process to tailor the shape shown in the upper inset of Fig. 1, which contaminated the stack boundary. The Ga$^+$ contamination resulted in a broad distribution of the switching current along the $c$ axis of the stack of $N$~80 junctions from 115 $\mu$A to a very low value of 0.2 $\mu$A at $T$=0.3 K [not shown] [19]. The four-terminal transport measurements were carried out in a He$^3$ cryostat with the base temperature of 0.28 K. To suppress the high-frequency noise, room-temperature π-filtering was used and the measurement leads were embedded in copper powder-filled epoxy and silver-paste glue at cryogenic temperatures. All the voltage measurements were done with battery powered pre-amplifiers (PAR113). The measured $c$-axis voltage represents the potential difference only across the gray-colored region in the lower inset of Fig. 1, because the rest parts of the junctions have much higher superconducting critical current. The microwave radiation was introduced to the sample through a semi-rigid coaxial cable and a copper-helix antenna, which was placed in front of the sample in a Faraday cage and was of the order of 10 mm in size.

## III. RESULTS AND DISCUSSION

Fig. 1 shows the *I-V* curves, measured without MW irradiation, in the low-current bias region at *T*=0.3 K. The so-called "quasiparticle" branches [1] are denoted by arrows. The noise-free critical current, $I_c$ at *T*=0.3 K was estimated to be 314 nA from the switching current distributions [20] and the corresponding plasma frequency $f_p = \sqrt{eI_c/\pi h C_J}$ was ~5 GHz, where the junction capacitance, $C_J$ = 1 pF, using the typical capacitance value per unit area of 45 fF/$\mu$m$^2$ for a single junction [21]. This value is much lower than the usual value of a few hundred GHz in the regular IJJs [7]. This extremely low plasma frequency conveniently allowed us to systematically study the MW irradiation frequency dependence of the Shapiro step behavior using few tens of GHz frequency-range.

Fig. 2 displays constant-voltage current steps in the stack of superconducting atomic layers (the Bi-2212 crystal), irradiated by MW with frequencies of 26 GHz and 13 GHz. The observed current steps are marked with the index numbers and denoted by the arrows. The index numbers correspond to the voltages which are integer multiples of $V_0 = hf/2e$, predicted by the AC Josephson relation [8,9,10]. In particular, the external frequencies of 26 GHz and 13 GHz correspond to the voltages of 53.76 $\mu$V and 26.88 $\mu$V, respectively. The voltage positions of the steps corresponding to *n*=1~6 in Fig. 2(a) are 53.67±4.5 $\mu$V, 108.03±5.8 $\mu$V, 158.42±4.7 $\mu$V, 213.37±3.4 $\mu$V, 267.42±3 $\mu$V, and 320.65±4.1 $\mu$V, respectively, which are integer multiples of $V_0$=53.76 $\mu$V at *f*=26 GHz. This excellent agreement between the theory and experiment confirms that these steps are genuine Shapiro steps. We observed the ZCSS for index numbers *n*=±1 (±1) and ±4 (±3) at *f*=26 (13) GHz. If $\nu$ is the number of junctions in resonance with the MW, the first steps should be observed at $\nu V_0$. The observations of the first current steps at *V*≈±54 $\mu$V and ±27 $\mu$V in figures of 2(a) and (b), respectively, imply that the Shapiro response occurred only in a single junction (*i.e.* $\nu=1$), although there were $N\sim80$ junctions in the stack. On the other hand, the current steps corresponding to the odd index numbers in the figures could have been the half integer Shapiro steps arising over two synchronized junctions. The widths of the resonant current steps at the odd index numbers, however, are comparable to those at the even index numbers. This indicates that the steps corresponding to the odd index numbers are not the subharmonic resonances but correspond to the main ($n = \pm1$) and harmonic resonances ($|n| > 1$).

The ZCSS observed in the IJJs of high-$T_c$ superconductors satisfy their observation condition expected based on the McCumber-Stewart RCSJ model [22]. The condition stipulates that clear Shapiro steps appear in the underdamped junctions when the impedance associated with the parallel capacitance is smaller than the lumped Josephson junction impedance associated with its kinetic inductance, $L_J$ [$=h/4\pi e I_c$]. The condition, equivalent to the relation $(f/f_p)^2 \gg 1$, ensures the linear voltage response to a sinusoidal current from the MW [17]. This requirement is already fulfilled with the frequency of 26 GHz (see the corresponding curves in Fig.2a). To form a ZCSS, the width of the *n*th current step, $\Delta I_n^{max} = I_C J_n^{max}$, should be larger than the bias current value corresponding to the end point of the step under consideration. This gives the second ZCSS-observing condition of $J_n^{max}/n > \Omega$ [$=f/f_c$], where $J_n^{max}$ is the maximum value of the *n*th order Bessel function, $f_c$ [$=2eV_c/h$] is the characteristic frequency of the junction, $V_c$ [$=I_c R_q$] is the characteristic voltage of the junction, and $R_q$ is the quasiparticle tunneling resistance in the subgap region [12]. The value of $R_q$ in the junction is estimated to be 22 k$\Omega$ using the *I-V* curves in Fig. 1 so that $V_c \sim 7$ mV. In the case of the $f$=26 GHz, the normalized frequency $\Omega$ is $\sim 7\times10^{-3}$ so that the second condition allows the observation of the ZCSS up to an order of n~30 order, which includes the case of *n*=4 as Fig. 2(a). The largest value of the step number at which ZCSS occurred in our samples was n=5.

Fig. 3 undoubtedly shows half-integer steps for *f*=18 GHz at various powers, as indicated by arrows at each curve (ZCSS is also visible but not marked by arrows). The inset of Fig. 3 also shows the quarter-integer steps such as *n*=25/4, 26/4, 29/4, and 30/4, where *n*=26/4 and 30/4 steps can also be half-integer steps. These fractional steps may be related to the nonsinusoidal current-phase relations due to the 2π-periodic odd-function character of supercurrent, $I_S(\phi)$, where $\phi$ is the phase difference between two superconducting electrodes. In the case of Bi-2212 IJJs, with *d*-wave order parameter, the higher harmonic terms cannot be neglected as the conventional *s*-wave type Josephson junctions [23], resulting in the nonsinusoidal current-phase relation. It is believed that the half-integer and quarter steps in Fig. 3 are from the components of $I_{c2}\sin 2\phi$ and $I_{c4}\sin 4\phi$ in the nonsinusoidal current-phase relation, respectively.

As the MW frequency was lowered, the ZCSS changed to non-zero-crossing Shapiro steps with hysteresis. Fig. 4(a) shows the *I-V* curves with *f*=5 GHz, a frequency close to $f_p$. At this frequency the widths of all observed steps reduced to a small fraction of $I_c$ [17]. Only the non-zero-crossing Shapiro steps were observed in the examined power range. It reconfirms that, although it satisfies the condition of $J_1^{\max} > \Omega$ [~1.3×10⁻³ at *f*=5 GHz] for *n*=1 [11,12,17], the irradiated MW frequency should be sufficiently larger than the junction plasma frequency ($(f/f_p)^2 \gg 1$) to observe the

ZCSS.

For frequencies well below $f_p$ nonlinear hysteretic curves were observed, but they did not exhibit Shapiro steps, as the ones for $f$=1 GHz in Fig. 4(b). The voltage value of the current step at $f$=1 GHz, denoted by arrows, increase with increasing MW irradiation power. Similar behavior was also observed for $f$=0.5 GHz [not shown]. This power dependence of the current step in the low-frequency range below $f_p$ is assumed to be caused by the fluxon motion induced by the *ac* magnetic field component of the MW [24]. *I-V* curves for *P*=-18.2 dBm in Fig. 4(b) also show zero-crossing steps near *V*=0.1 mV but their mechanism is not related to the phase synchronization phenomenon [25]. This hysteretic region near the zero bias for *P*=-18.2 dBm evolves into two hysteretic regions for *P*=-14.7 dBm. These two regions denoted by gray color in Fig. 4(b) correspond to the gray-filled regions in Fig. 4(a). For $f$=5 GHz, a sudden jump to a higher voltage state in the hysteretic region caused the disappearance of the Shapiro steps for $n=\pm 5$-$\pm 12$. In Fig. 2(b) with $f$=13 GHz, the absence of steps for n=±7-±12 also occurs in the hysteretic region. In this frequency ranges near $f_p$, the steps occur with this hysteresis as a background. The effect on the Shapiro steps due to the hysteresis weakens with increasing MW frequency and the well-developed steps ($n=\pm 2, \pm 3$) in this region were observed for $f$=26 GHz as in Fig. 2(a). These results are consistent with frequency condition of $f \gg f_p$ to observe the stable Shapiro steps in a JJ [11,12].

On the other hand, the finite resistive slopes seen in the voltage plateaus in Fig. 4(a) were caused by the thermal fluctuation, with very small resonating current-step width [16]. If the thermal energy $k_B T$ ($k_B$; the Boltzmann constant) is comparable to the stored energy in the resonant current, $\hbar \Delta I/2e$ ($\Delta I$; the width of the current step) [10], the corresponding Shapiro step is strongly affected by the thermal fluctuation and exhibits a finite resistive slope. To figure out this temperature effect, we observed the temperature dependence of the step slope with a fixed frequency ($f$=23 GHz) and a power ($P$=-17 dBm) as Fig. 5. With increasing temperature, the shape of steps got smeared and acquires a larger differential resistance, while the current-step widths were suppressed. The thermal energy at $T$=0.6 K corresponds to the current fluctuation estimate of $\Delta I_{th}$=25 nA, based on the relation of $k_B T=\hbar \Delta I_{th}/2e$. The current steps in the condition of $\Delta I \leq \Delta I_{th}$ will be strongly affected by the thermal noise, resulting in the finite slope in the current steps. The current steps with $\Delta I$=4.4 nA at n=±3, which is less than $\Delta I_{th}$, indeed showed the resistance of 2.8 kΩ at $T$=0.6 K and disappeared at $T$~2K. The current steps of $n$=±5 with $R$=100 Ω and $\Delta I$=48 nA at $T$=0.6 K also evolved to that with R=2 kΩ and $\Delta I$=11nA at $T$=3.53 K. The observed suppression of $\Delta I$ with increasing temperature is due to the premature switching by the thermal noise based on the Stewart-McCumber washboard potential model [22]. The increase of the resistance in the current steps with increasing temperature originates from the enhancement of the phase diffusion effect [16].

## IV. SUMMARY

In summary, the junction plasma frequency was conveniently tuned by the FIB irradiation of high-$T_c$ crystals. The corresponding low plasma frequency and the switching current, which was in the range of a few hundred nA, allowed us to observe ZCSS from a single junction in the MW frequency range of 10 GHz. This rectification effect consists of the observation of a quantized, frequency-proportional dc voltage on the junctions when only the microwave irradiation is applied, and the dc bias current being zero. In this study we also reconfirmed two conditions of $J_n^{max}/n > \Omega$ and $(f/f_p)^2 >> 1$ being the requirements for the occurrence of this rectification phenomenon. This system showed fractional Shapiro steps, which suggests a non-sinusoidal character of the current-phase relation in the IJJ system in Bi-2212 high-$T_c$ superconductors. The finite slope of the steps is due to thermal fluctuations, which is related to the phase diffusion phenomena. In the application point of view, the reported high-intensity FIB tailoring should provide a convenient approach to the practical voltage-standard applications of Bi-2212 single crystals, also.

## ACKNOWLEDGEMENTS

This work was supported by DOE Grants No. DEFG02-07ER46453 and No. DEFG02-91ER45439. We acknowledge the access to the fabrication facilities at the Frederick Seitz Materials Research Laboratory. This work was also supported by POSTECH Core Research Program and the Korea Research Foundation Grants No. KRF-2006-352-C0020.

**FIGURE CAPTIONS**

Figure 1. *I-V* curves of the Bi-2212 layered superconducting crystal, without microwave irradiation at *T*=0.3 K. The arrows indicate the quasiparticle branches. Upper inset: an SEM micrograph of the sample, where the scale bar is 5 $\mu$m. Lower inset: schematic geometry of the sample and the measurement configuration. The gray region on the middle, marked with dashed lines, represents the volume of the sample in which the applied current flows perpendicular to the layers of the Bi-2212 superconductor.

Figure 2. Microwave response of the Bi-2122 layered superconductor to external microwave radiation at *T*=0.3 K. The dc voltage on the sample is plotted versus the dc bias current. (a) The radiation frequency is *f*=26 GHz and the nominal radiation power is *P*=-12.3 dBm (b) The frequency and the power are *f*=13 GHz with *P*=-2.9 dBm. The numbers are defined as *n=2eV/hf* and represent the ratio of the frequency of the phase difference-temporal oscillation and the applied microwave signal frequency.

Figure 3. (color online) Microwave response of the IJJs stack for *f*=18 GHz with two different powers at *T*=0.3 K. The voltage axis was normalized by the voltage corresponding to the external frequency. The arrows indicate the half-integer steps at each power. The inset shows the quarter-integer steps as indicated by arrows and the fractional index numbers at two different powers at *f*=18 GHz.

Figure 4. (color online) Microwave response of the IJJs stack for (a) *f*=5 GHz with the irradiation power of P=-22.4 dBm and for (b) *f*=1 GHz with *P*=-14.7 dBm and -18.2 dBm at *T*=0.3 K. The index numbers and arrows in (a) are defined in the same manner

as in Fig. 2. The arrows in (b) indicate the positions of the nearly constant-voltage current steps.

Figure 5. (color online) Temperature dependence of Shapiro steps with $f$=23 GHz and $P$=-17 dBm at $T$= 0.6 K , 1 K , 1.4 K, 2.28 K, and 3.53 K, respectively, from the left most one. The data were shifted horizontally for the clarity as 30 nA.

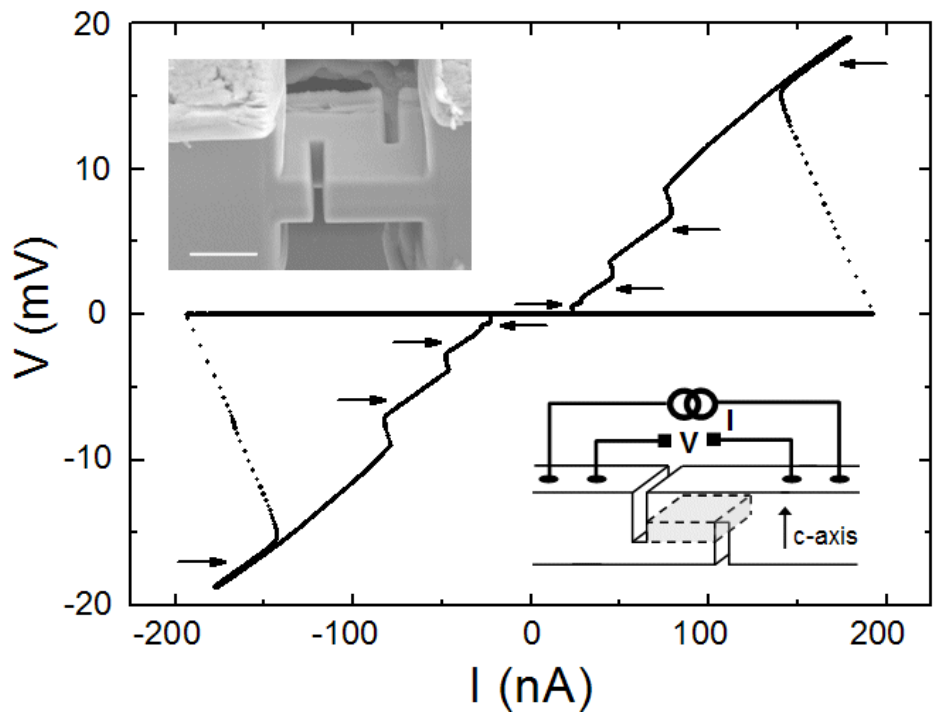

Fig. 1

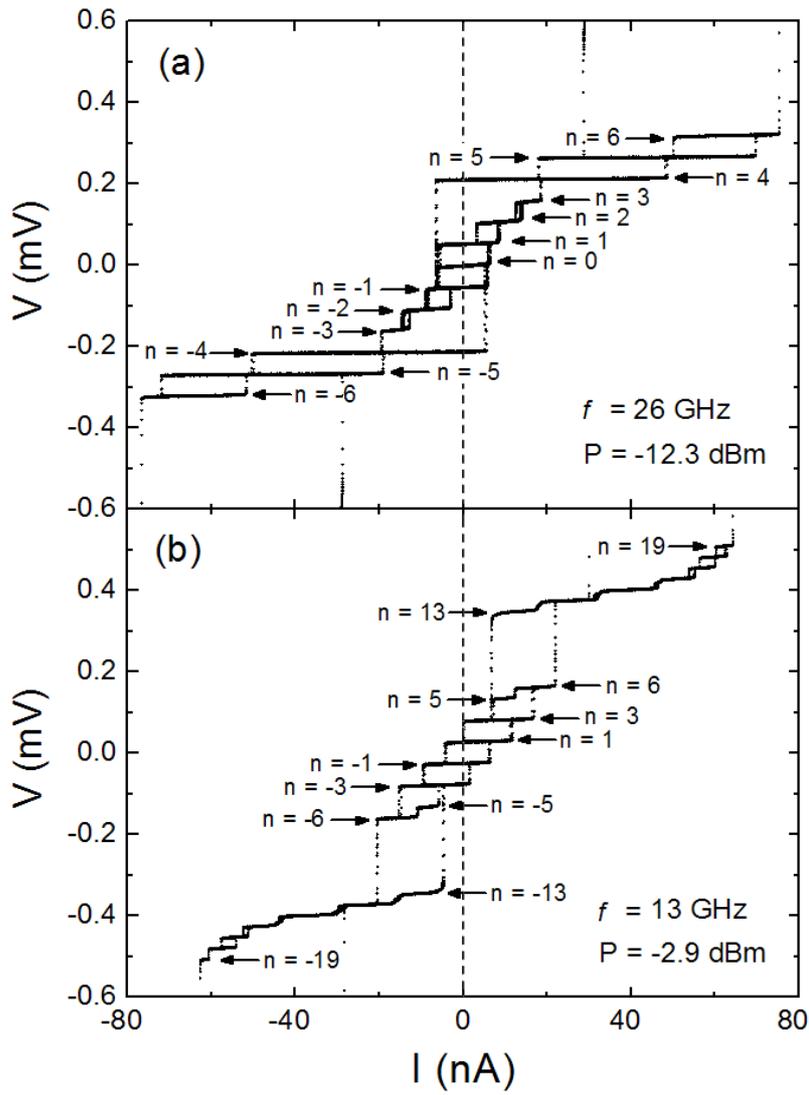

Fig. 2

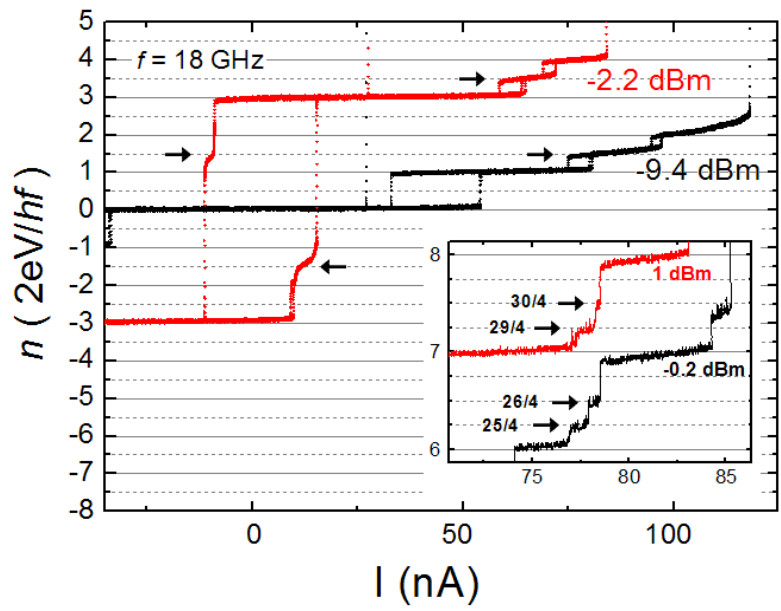

Fig. 3

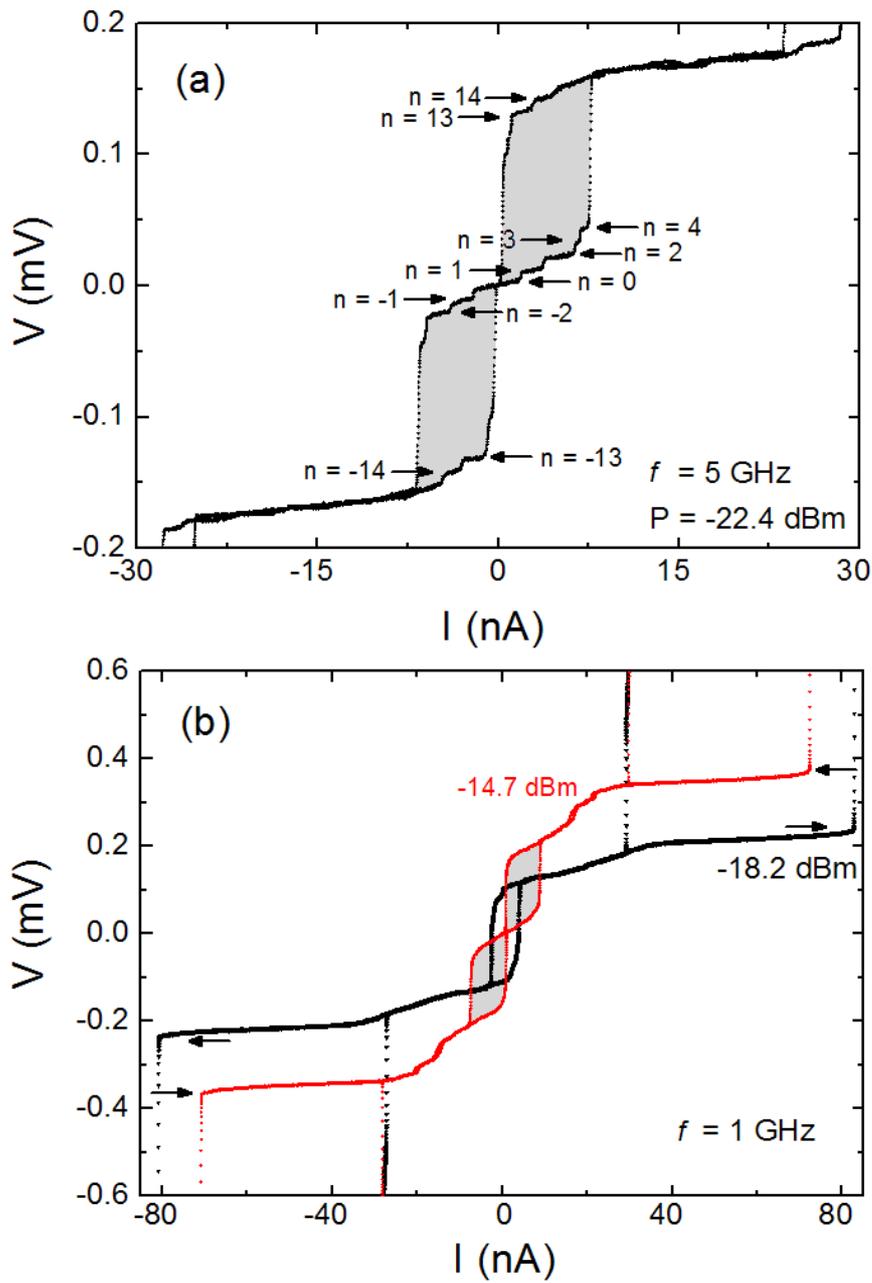

Fig. 4

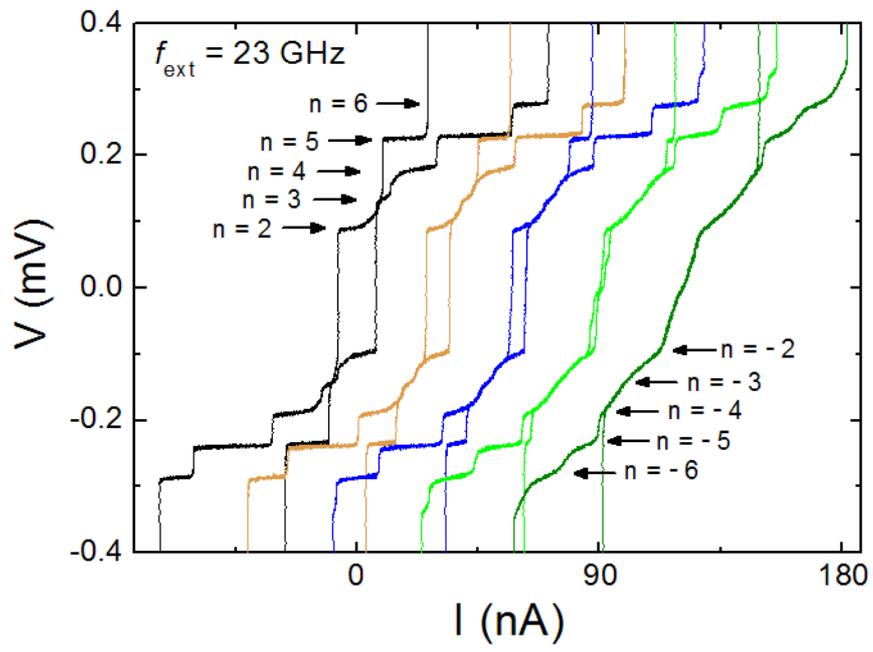

Fig. 5